\documentclass[conference]{IEEEtran}

\usepackage{verbatim}

\ifCLASSINFOpdf
\else
\fi
\usepackage{algorithm}
\usepackage{algorithmic}
\usepackage{graphicx}
\usepackage{amsmath}
\usepackage{amsfonts}
\usepackage{amssymb}
\usepackage{algorithm}
\usepackage{algorithmic}

\newcommand{\beq}{\begin{equation}}
\newcommand{\eeq}{\end{equation}}
\newcommand{\beqnn}{\begin{equation*}}
\newcommand{\eeqnn}{\end{equation*}}
\newcommand{\beqy}{\begin{eqnarray}}
\newcommand{\eeqy}{\end{eqnarray}}
\newcommand{\beqynn}{\begin{eqnarray*}}
\newcommand{\eeqynn}{\end{eqnarray*}}
\newcommand{\bit}{\begin{itemize}}
\newcommand{\eit}{\end{itemize}}
\newcommand{\ben}{\begin{enumerate}}
\newcommand{\een}{\end{enumerate}}

\newcommand{\balg}[1]{\begin{algorithm} \caption{#1}}
\newcommand{\ealg}{\end{algorithm}}

\newcommand{\balgc}{\begin{algorithmic}[1]}
\newcommand{\ealgc}{\end{algorithmic}}

\newcommand{\bary}{\begin{array}}
\newcommand{\eary}{\end{array}}
\newcommand{\bmx}{\begin{bmatrix}}
\newcommand{\emx}{\end{bmatrix}}
\newcommand{\bsmx}{\left[\begin{smallmatrix}}
\newcommand{\esmx}{\end{smallmatrix}\right]}
\newcommand{\bmxc}[1]{\left[\begin{array}{@{}#1@{}}}
\newcommand{\emxc}{\end{array}\right]}
\newcommand{\bcn}{\begin{center}}
\newcommand{\ecn}{\end{center}}

\newcommand{\nbn}{{n \times n}}

\newcommand{\cnd}{\kappa}

\newcommand{\D}{\boldsymbol{D}}

\newcommand{\G}{\boldsymbol{G}}
\renewcommand{\H}{\boldsymbol{H}}
\newcommand{\I}{\boldsymbol{I}}

\renewcommand{\P}{\boldsymbol{P}}
\newcommand{\Q}{\boldsymbol{Q}}
\newcommand{\R}{\boldsymbol{R}}

\newcommand{\U}{\boldsymbol{U}}
\newcommand{\V}{\boldsymbol{V}}

\newcommand{\Z}{\boldsymbol{Z}}

\newcommand{\e}{\boldsymbol{e}}

\renewcommand{\l}{\boldsymbol{l}}

\newcommand{\rr}{\boldsymbol{r}}

\renewcommand{\v}{\boldsymbol{v}}

\newcommand{\x}{{\boldsymbol{x}}}
\newcommand{\y}{{\boldsymbol{y}}}
\newcommand{\z}{\boldsymbol{z}}

\newcommand{\by}{{\bar{y}}}
\newcommand{\bz}{{\bar{z}}}
\newcommand{\bc}{{\bar{c}}}
\newcommand{\br}{{\bar{r}}}

\newcommand{\bbz}{{\bar{\z}}}
\newcommand{\bby}{{\bar{\y}}}

\newcommand{\bbR}{{\bar{\R}}}

\providecommand{\norm}[1]{\lVert#1\rVert}

\providecommand{\rnd}[1]{\lfloor #1 \rceil}

\begin{document}
%
\title{Partial LLL Reduction}

\author{\IEEEauthorblockN{Xiaohu Xie}
\IEEEauthorblockA{School of Computer Science\\
McGill University\\
Montreal, Quebec, Canada H3A 2A7\\
Email: xiaohu.xie@mail.mcgill.ca}
\and
\IEEEauthorblockN{Xiao-Wen Chang}
\IEEEauthorblockA{School of Computer Science\\
McGill University\\
Montreal, Quebec, Canada H3A 2A7\\
Email: chang@cs.mcgill.ca}
\and\IEEEauthorblockN{Mazen Al Borno}
\IEEEauthorblockA{Department of Computer Science\\
University of Toronto\\
Toronto, Ontario, Canada M5S 2E4\\
Email: mazen@dgp.toronto.edu}
}


%


\maketitle

\begin{abstract}
The Lenstra-Lenstra-Lovasz (LLL) reduction has  wide applications in digital communications.
It can greatly improve the speed of the sphere decoding (SD) algorithms for solving an integer least squares (ILS) 
problem and  the performance of the Babai integer point, a suboptimal solution to the ILS problem. 
Recently Ling and Howgrave-Graham proposed the so-called effective LLL (ELLL) reduction.
It has less computational complexity than  LLL, 
while it has the same effect on the performance of the Babai integer point as LLL.
In this paper we propose a partial LLL (PLLL) reduction.
PLLL avoids the numerical stability problem with ELLL, which may result in very poor
performance of the Babai integer point.
Furthermore, numerical simulations indicated that it is faster than ELLL.
We also show that in theory PLLL and ELLL have the same effect  on
the search speed of a typical SD algorithm as LLL.
\end{abstract}


%
\IEEEpeerreviewmaketitle

\section{Introduction}
In a multiple-input and multiple-output (MIMO) system, often we have the following linear model:
\beq
\label{e:model}
\y=\H\x+\v,
\eeq
where $\y\in \mathbb{R}^n$ is the channel output vector,  $\v\in \mathbb{R}^n$ is the noise vector following a normal distribution
$\mathcal{N}(\boldsymbol{0},\sigma^2 \I)$, $\H\in\mathbb{R}^{n\times m}$ is the channel matrix, and $\x\in\mathbb{Z}^m$ is the unknown integer data vector. 
In some applications where complex ILS problems may need to be solved instead, we can first transform the complex ILS problems to equivalent real ILS problems. 
For simplicity, like \cite{LinH07}, in this paper we assume $m=n$ and $\H$ is nonsingular. 

To estimate $\x$, one solves an integer least squares problem
\beq
\label{e:ILS}
\min_{\x\in{Z}^n}\|\y-\H\x\|_2^2,
\eeq
which gives  the maximum-likelihood estimate of $\x$. 
It has been proved that the ILS problem is NP-hard \cite{Boa81}. 
For applications which have high real-time requirement, an approximate solution of \eqref{e:ILS} is usually computed instead. 
A often used approximation method is the nearest plane algorithm proposed by Babai \cite{Bab86}
and the produced approximate integer solution is referred to as the Babai integer point.
In communications, a method for finding this approximate solution is referred to as 
a successive interference cancellation decoder.

A typical method to solve \eqref{e:ILS} is a sphere decoding (SD) algorithm, such as
the Schnorr-Euchner algorithm  (see \cite{SchE94} and  \cite{AgrEVZ02})
or its variants (see, e.g., \cite{DamGC03} and \cite{ChaH05}).
A  SD algorithm has two phases.
First the reduction phase  transforms  \eqref{e:ILS} to an equivalent problem.
Then the search phase enumerates integer points in a hyper-ellipsoid 
to find the optimal solution.
The   reduction phase  makes the search phase easier and more efficient.
The  Lenstra-Lenstra-Lovasz (LLL) reduction \cite{LenLL82} is the mostly used reduction in practice.
An LLL reduced basis matrix has to satisfy two conditions.
One is the size-reduction condition and the other is the Lovasz condition
(see Section \ref{s:lll} for more details).
Recently  Ling and Howgrave-Graham \cite{LinH07} argued geometrically that the size-reduction condition 
does not change the performance of the Babai integer point. 
Then they proposed the so-called  effective LLL reduction (to be referred to as ELLL) which mostly avoids 
size reduction. 
They proved that their ELLL algorithm has less time complexity than the original LLL algorithm
given in \cite{LenLL82}.
However, as implicitly pointed out in \cite{LinH07}, the ELLL algorithm has a numerical stability problem.
Our simulations, presented in Section \ref{s:sim}, will indicate that 
ELLL may give a very bad estimate of $\x$ than the LLL reduction due to its numerical stability problem.

In this paper, we first show algebraically that the size-reduction condition of the LLL reduction 
has no effect on   a typical SD search process.
Thus it has no effect on the performance of the Babai integer point, the first integer point found in the search process.
Then we propose a partial LLL reduction algorithm, to be referred to as PLLL, 
which avoids the numerical stability problem
with ELLL.  Numerical simulations indicate that it is faster than ELLL
and is  as numerically stable as LLL.

\section{LLL Reduction}\label{s:lll}
In matrix language, the LLL reduction can be described as a QRZ factorization \cite{ChaG09}:
\beq
\label{e:QRZ}
\Q^T \H \Z = \R,
\eeq
where $\Q  \in \mathbb{R}^{n\times n}$ is orthogonal,
$\Z\in   \mathbb{Z}^{n\times n}$ is a  unimodular matrix (i.e., $\det(\Z)=\pm1$),
and  $\R\in \mathbb{R}^{n\times n}$ is upper triangular and satisfies the following two conditions:
\beq \label{e:lll}
\begin{split}
& | r_{i,j}| \le | r_{i,i}|/2,  \ \  1\leq i<j \leq n  \\
& \delta\,   r_{i-1,i-1}^2  \le    r^2_{i-1,i}+r^2_{i,i}, \ \ 1<i \leq n,
\end{split}
\eeq
where the parameter $\delta\in (1/4, 1]$.
The first condition in \eqref{e:lll} is the size-reduction condition
and the second condition in  \eqref{e:lll} is the Lovasz condition.

Define $\bby=\Q^T\y$ and $\z=\Z^{-1}\x$.
Then it is easy to see that the ILS problem \eqref{e:ILS} is reduced to 
\beq \label{e:ILSR}
\min_{\z\in \mathbb{Z}^n}\|\bby-\R\z\|^2_2.
\eeq
If $\hat{\z}$ is the solution of the reduced ILS problem \eqref{e:ILSR}, then $\hat{\x}=\Z\hat{\z}$
is the ILS solution of the original problem \eqref{e:ILS}.

The LLL algorithm first applies the Gram-Schmidt orthogonalization (GSO) to $\H$, 
finding the QR factors  $\Q$   and $\R$
(more precisely speaking, to avoid square root computation, the original LLL algorithm 
gives a column scaled $\Q$ and a row scaled $\R$ which has unit   diagonal entries). 
Two types of basic unimodular matrices are then implicitly used to update $\R$ 
so that it satisfies   \eqref{e:lll}:
integer Gauss transformations (IGT) matrices and  permutation matrices,
see below.


To meet the first condition in \eqref{e:lll}, we can apply an IGT:
\beqnn
\Z_{ij}=\I-\zeta \e_i\e_j^T.
\eeqnn
where $\e_i$ is the $i$-th column of $\I_n$. It is easy to verify that $\Z_{ij}$ is unimodular. Applying $\Z_{ij}\ (i < j)$ to $\R$ from the right gives
\beqnn
\bbR= \R\Z_{ij} = \R-\zeta \R\e_i\e_j^T,
\eeqnn
Thus $\bbR$ is the same as $\R$, except that $\br_{kj} = r_{kj} - \zeta r_{ki},\ k = 1, \ldots, i$. By setting $\zeta= \lfloor r_{ij}/r_{ii} \rceil $, 
the nearest integer to $r_{ij}/r_{ii}$, we ensure $|\br_{ij}|\le|\br_{ii}|/2$. 


To meet the second condition in \eqref{e:lll}, we permutations columns. 
Suppose that we interchange columns $i-1$ and $i$ of $\R$.
Then  the upper triangular structure of $\R$ is no longer maintained. 
But we can bring $\R$ back to an upper triangular matrix by using the GSO technique
(see \cite{LenLL82}): 
\beqnn
\label{e:Permu}
\bbR=\G_{i-1,i}\R\P_{i-1,i},
\eeqnn
where $\G_{i-1,i}$ is  an orthogonal matrix and $\P_{i-1,i}$ is a permutation matrix.
Thus,
\begin{align}
\br_{i-1,i-1}^2&= r^2_{i-1,i}+r^2_{i,i},\label{e:PerR}\\
\br^2_{i-1,i}+\br^2_{i,i}&= r_{i-1,i-1}^2. \nonumber
\end{align}
If $\delta\, r_{i-1,i-1}^2 > r^2_{i-1,i}+r^2_{i,i}$, then the above operation 
guarantees $\delta\, \br_{i-1,i-1}^2 <  \br^2_{i-1,i}+\br^2_{i,i}$.

The LLL reduction process is described in Algorithm \ref{a:LLL}.

\begin{algorithm}
\caption{LLL reduction}   \label{a:LLL}

\begin{algorithmic}[1]
  \STATE apply GSO to obtain $\H=\Q\R$;
  \STATE set $\Z=\I_n$, $k=2$;
  \WHILE{$k\le n$}
   \STATE apply IGT $\Z_{k-1,k}$ to reduce $r_{k-1,k}$: $\R=\R\Z_{k-1,k}$;
   \STATE update $\Z$: $\Z=\Z\Z_{k-1,k}$;
   \IF{$\delta\,  r_{k-1,k-1}^2> \left(r^2_{k-1,k}+r^2_{k,k}\right)$}
    \STATE permute and triangularize $\R$: $\R\!=\!\G_{k-1,k}\R\P_{k-1,k}$;\label{l:pt}
    \STATE update $\Z$: $\Z=\Z\P_{k-1,k}$;
    \STATE $k=k-1$, when $k>2$;
   \ELSE
    \FOR{$i=k-2,\dots,1$}	\label{LLL:elll1}
      \STATE apply IGT $\Z_{ik}$ to reduce $r_{ik}$: $\R=\R\Z_{ik}$; 
      \STATE update $\Z$: $\Z=\Z\Z_{i,k}$;
    \ENDFOR \label{LLL:elll2}
    \STATE $k=k+1$;
   \ENDIF
  \ENDWHILE
\end{algorithmic}
\end{algorithm}


\section{SD Search Process and Babai Integer Point}
For later use we  briefly introduce 
the often used SD search process (see, e.g.,  \cite[Section II.B.]{ChaH05}), which is a depth-first search (DFS)
through a tree.
The idea of SD is to search for the optimal solution of \eqref{e:ILSR} in a hyper-ellipsoid defined as follow:
\beq
\label{e:SD}
\|\bby-\R\z\|_2^2< \beta.
\eeq
Define  
\beq
\label{e:CK}
\begin{split}
& c_n = \by_n/r_{nn},\\
& c_k =\big (\by_k-\sum_{j=k+1}^n r_{kj}z_j \big)/r_{kk}, \ \ k=n-1,\ldots,1.
\end{split}
\eeq
Then it is easy to show that \eqref{e:SD} is equivalent to 
\beq
\label{e:SDLEVEL}
\mbox{level } k:  \qquad r_{kk}^2(z_k-c_k)^2<\beta -\sum_{j=k+1}^n r_{jj}^2(z_j-c_j)^2, 
\eeq
where $k=n, n-1,\dots,1$.

Suppose $z_n,  z_{n-1}, \ldots, z_{k+1}$ have been fixed, we try to determine $z_k$ at level $k$
by using   \eqref{e:SDLEVEL}. 
We first compute $c_k$ and then take $z_k=\rnd{c_k}$.
If  \eqref{e:SDLEVEL} holds, we move to level $k-1$ to try to fix $z_{k-1}$.
If at level $k-1$, we cannot find any integer for $z_{k-1}$ such that  
 \eqref{e:SDLEVEL}  (with $k$ replaced by $k-1$) holds, we move back to level $k$
and take $z_k$ to be the next nearest integer to $c_k$.
If  \eqref{e:SDLEVEL} holds for the chosen value of $z_k$, we again move to level $k-1$;
otherwise we move back to level $k+1$, and so on.
Thus  after   $z_n,\dots,z_{k+1}$ are fixed, 
we  try all possible values of $z_k$ in the following order until  
\eqref{e:SDLEVEL} dose not hold anymore
and we move back to level $k+1$:
\beq
\label{e:ORDER}
\begin{split}
& \rnd{c_k},\rnd{c_k}-1,\rnd{c_k}+1,\rnd{c_k}-2,\dots, ~\text{if } c_k\le\rnd{c_k}, \\
& \rnd{c_k},\rnd{c_k}+1,\rnd{c_k}-1,\rnd{c_k}+2,\dots, ~\text{if }c_k>\rnd{c_k}.
\end{split}
\eeq
When we reach level 1, we compute $c_1$ and take $z_1=\rnd{c_1}$.
If \eqref{e:SDLEVEL} (with $k=1$) holds, an integer point, say $\hat{\z}$, is found.
We update $\beta$ by setting $\beta=\|\y-\R\hat{\z}\|_2^2$
and  try to update $\hat{\z}$ to find a better integer point in the new hyper-ellipsoid.
Finally when we cannot find any new value for $z_n$ at level $n$ such that the corresponding 
inequality holds,  the search process stops and the latest found integer point is the optimal solution we seek.

At the beginning of the search process, we set $\beta=\infty$. 
The first integer point $\z$ found in the search process is referred to as the Babai integer point.

\section{Partial LLL Reduction}
\subsection{Effects of size reduction on search}
Ling and Howgrave-Graham \cite{LinH07} has argued geometrically that the performance of the Babai integer point 
is not affected by size reduction (see the first condition in \eqref{e:lll}). 
This result can be extended. 
In fact we will prove algebraically that the search process is not affected by size reduction.

We stated in Section \ref{s:lll}  that the size-reduction condition in \eqref{e:lll} is met by using IGTs.
It will be sufficient if we can show that one IGT will not affect the search process.
Suppose that two upper triangular matrices $\R\in \mathbb{R}^{n\times n}$ and $\bbR\in \mathbb{R}^{n\times n}$ 
have the relation:
$$
\bbR=\R\Z_{st}, \ \ \Z_{st}=\I-\zeta \e_s \e_t^T, \ \ s<t.
$$
Thus, 
\begin{alignat}{2}
\br_{kt}&=  r_{kt}-\zeta r_{ks},  \qquad &  \text{if } &  k \le s,   \label{e:IGTIne} \\
\br_{kj}&=  r_{kj},   &  \text{if } &  k >s \text{ or } j \neq t. \label{e:IGTEqu}
\end{alignat}
Let $\bbz = \Z_{st}^{-1}\z$.
Then the ILS problem \eqref{e:ILSR} is equivalent to 
\beq
\min_{\bbz \in \mathbb{Z}^n} \|\bby-\bbR \bbz\|_2^2.
\label{e:BILSR}
\eeq
For this ILS problem, the inequality the search process needs  to check at level $k$ is
\beq
\label{e:bSDLEVEL}
\mbox{level } k: \ \ \br_{kk}^2(\bz_k-\bc_k)^2<\beta -\sum_{j=k+1}^n \br_{jj}^2(\bz_j-\bc_j)^2,
\eeq 
Now we look at the search process for the two equivalent ILS problems.

Suppose  $\bz_n, \bz_{n-1}, \ldots, \bz_{k+1}$ and $z_n, z_{n-1}, \ldots, z_{k+1}$  have been fixed.
We consider the search process at level $k$  under three different cases.
\bit
\item Case 1: $k > s$.
Note that $\bbR_{k:n,k:n}=\R_{k:n,k:n}$. 
It is easy to see that we must  have 
$\bc_i=c_i$ and $\bz_i=z_i$ for $i=n, n-1, \ldots, k+1$.
Thus at level $k$, $\bc_k=c_k$ and the search process takes an identical value for $\bz_k$ and $z_k$.
For the chosen value, the two inequalities \eqref{e:SDLEVEL} and \eqref{e:bSDLEVEL}
are identical. So both hold or fail at the same time.


\item Case 2: $k=s$. 
According to Case 1, we have $\bz_{i}=z_i$ for $i=n,n-1,\ldots,s+1$. Thus
\begin{align*}
\bc_k&= \frac{\by_k-\sum_{j=k+1}^n \br_{kj}\bz_j}{\br_{kk}}\\
&= \frac{\by_k-\sum_{j=k+1,j\ne t}^n r_{kj}z_j-(r_{kt}-\zeta r_{kk})z_t}{r_{kk}}\\
&= c_k+\zeta z_t,
\end{align*}
where $\zeta$ and $z_t$ are integers. 
Note that $z_k$ and $\bz_k$ take on values according to \eqref{e:ORDER}. 
Thus values of $z_k$ and $\bz_k$ taken by the search process at level $k$ must
satisfy $\bz_k=z_k+\zeta z_t$. 
In other words, there exists one-to-one mapping between the values of $z_k$ and $\bz_k$.
For the chosen values of $\bz_k$ and $z_k$,  $\bz_k-\bc_k=z_k-c_k$.
Thus, again the two inequalities \eqref{e:SDLEVEL} and \eqref{e:bSDLEVEL} 
are identical. Therefore both inequalities hold or fail at the same time.

\item Case 3: $k <s$. 
According to Case 1 and Case 2,   $\bz_i=z_i$ for $i=n,n-1,\ldots,s+1$ and $\bz_s=z_s+\zeta z_t$.
Then for $k=s-1$,
\begin{align*}
\bc_{k}&=  \frac{\by_{k}-\sum_{j=k+1}^n \br_{kj}\bz_j}{\br_{kk}} \\
&= \frac{\by_k-\sum_{j=k+2,j\ne t}^n r_{kj}z_j-r_{ks}\bz_s-\br_{kt}z_t}{r_{kk}}\\
&= \frac{\by_k-\sum_{j=k+1}^n r_{kj}z_j-\zeta r_{ks}z_t+\zeta r_{ks}z_t}{r_{kk}}\\
&= c_{k}.
\end{align*}
Thus the search process takes an identical value for $\bz_k$ and $z_k$ when $k=s-1$. 
By induction we can similarly show this is true for a general $k<s$.
Thus, again the two inequalities \eqref{e:SDLEVEL} and \eqref{e:bSDLEVEL} 
are identical. Therefore they  hold or fail at the same time.
\eit

In the above we have proved that the search process is identical for both ILS problems 
\eqref{e:ILSR} and \eqref{e:BILSR} (actually the two search trees have an identical structure).
Thus the speed of the search process  is not affected by  the size-reduction condition in \eqref{e:lll}.
For any two integer points $\bbz^\ast$ and $\z^\ast$ found in the search process at the same time
for the two ILS problems,  
we have seen that $\bz_i^\ast=z_i^\ast$ for $i=n, \ldots, s+1, s-1, \ldots, 1$ 
and $\bz_s^\ast=z_s^\ast+\zeta z_t^\ast$, i.e.,   $\bbz^\ast=\Z_{st}^{-1}\z^\ast$. 
Then
$$
\|\bby-\bbR\bbz^\ast\|_2^2 = \|\bby-\R\z^\ast\|_2^2.
$$
Thus, the performance of the Babai point is not affected by the size-reduction condition in \eqref{e:lll} either,
as what \cite{LinH07} has proved from a geometric perspective.

However, the IGTs which reduce the super-diagonal entries of $\R$ are not useless 
when they are followed by permutations. 
Suppose $|r_{i-1,i}|>\frac{|r_{i-1,i-1}|}{2}$.
If we apply $\Z_{i-1,i}$ to reduce $r_{i-1,i}$, permute columns $i-1$ and $i$ of $\R$
and triangularize it, we have from \eqref{e:PerR} and \eqref{e:IGTIne} that 
\begin{align*}
\br_{i-1,i-1}^2&= \left(r_{i-1,i}-\left\lfloor\frac{r_{i-1,i}}{r_{i-1,i-1}}\right\rceil r_{i-1,i-1}\right)^2+r^2_{ii}\\
& < r_{i-1,i}^2+r^2_{ii}.
\end{align*}
From \eqref{e:PerR} we observe that the IGT can make $|r_{i-1,i-1}|$ smaller 
after permutation and triangularization.
Correspondingly $|r_{i,i}|$ becomes larger, as it is easy to prove that
$|r_{i-1,i-1}r_{i,i}|$ remains unchanged  after the above operations.  

The ELLL algorithm given in \cite{LinH07}  is essentially identical to Algorithm \ref{a:LLL} 
after lines \ref{LLL:elll1}--\ref{LLL:elll2}, which reduce other off-diagonal entries of $\R$, are removed.

\subsection{Numerical stability issue}
We have shown that in the LLL reduction, an IGT is useful only if it reduces a super-diagonal entry.
Theoretically, all other IGTs will have no effect on the search process. 
But simply removing those IGTs can causes serious numerical stability problem even $\H$ is not ill conditioned. 
The main cause of the stability problem is that during the reduction process, some entries of $\R$ 
may grow significantly. 
For the following $n\times n$ upper triangular matrix 
\beq \label{e:unstableA}
\H=
\bmx
1   &2   &4   &    &      & \\
    &1   &2   &0   &      & \\
    &    &1   &2   &4     & \\
    &    &    &1   &2     & \ddots\\
    &    &    &    &1     & \ddots\\
    &    &    &    &      & \ddots\\ 
\emx,
\eeq
when $n=100$, the condition number $\cnd_2(\H) \approx 34$. 
The LLL reduction will reduce $\H$ to an identity matrix $\I$. 
However, if we apply the ELLL reduction, the maximum absolute value in $\R$ will be $2^{n-1}$.
When $n$ is big enough, an integer overflow will occur.

In the ELLL algorithm, the super-diagonal entries are always reduced.
But if a permutation does not occur immediately after the size reduction, then this  size reduction is useless
in theory and furthermore it may help the growth of the other off-diagonal entries in the same column.
Therefore, for efficiency and numerical stability, 
we propose a  new strategy of applying IGTs in Algorithm \ref{a:LLL}. 
First we compute $\zeta=\rnd{r_{k-1,k}/r_{k-1,k-1}}$. 
Then we test  if  the following inequality 
$$
\delta\, r^2_{k-1,k-1} > \left(r_{k-1,k}-\zeta r_{k-1,k-1}\right)^2 + r^2_{kk} 
$$
holds. 
If it does not, then the permutation of columns $k-1$ and $k$  will not occur, 
no IGT will be applied, and the algorithm moves to column $k+1$.
Otherwise, if $\zeta \neq 0$, the algorithm reduces $r_{k-1,k}$
and if $|\zeta|\geq 2$, the algorithm also reduces all $r_{i,k}$ for $i=k-2, k-3, \dots,1$ for stability consideration. 
When $|\zeta|=1$, we did not notice any stability problem if we do not reduce the above size 
of $r_{i,k}$ for $i=k-2, k-3, \dots,1$.

\subsection{Householder QR with minimum column pivoting}
In the original LLL reduction and the ELLL reduction, GSO is used to compute the QR factorization of $\H$ 
and to update $\R$ in the later steps. 
The cost of computing the QR factorization by GSO is  $2n^3$ flops,
larger than $4n^3/3$ flops required by the QR factorization by Householder reflections
(note that we do not need to form the $Q$ factor explicitly in the reduction process);
see, e.g., \cite[Chap 5]{GolV96}.
Thus we propose to compute the QR factorization  by Householder  reflections instead of  GSO.

Roughly speaking, the reduction would like to have small diagonal entries at the beginning
and large diagonal entries at the end. 
In our new reduction algorithm, the IGTs are applied only when a permutation will occur. 
The less occurrences of permutations, the faster the new reduction algorithm runs. 
To reduce the occurrences of permutations in the reduction process, 
we propose to compute the QR factorization with minimum-column-pivoting:
\beq
\Q^T \H\P = \R
\eeq
where $\P\in\mathbb{Z}^\nbn$ is a permutation matrix.  
In the $k$-th step of the QR factorization, we find the column in $\H_{k:n,k:n}$, say column  $j$,
which has the minimum 2-norm.  
Then we interchange columns $k$ and $j$ of $\H$.  
After this we do what the $k$-th step of a regular Householder QR factorization does.
Algorithm \ref{a:PQR} describes the process of the factorization.

\begin{algorithm}
 \caption{QR with minimum-column-pivoting}\label{a:PQR}



\begin{algorithmic}[1]
  \STATE set $\R=\H,\P=\I_n$;
  \STATE compute $l_k=\norm{\rr_k}_2^2$, $k=1\dots,n$;
  \FOR{$k=1, 2, \ldots, n$}
    \STATE find $j$ such that $l_j$ is the minimum among $l_k,\dots,l_n$;
    \STATE exchange columns $k$ and $j$ of $\R$, $\l$ and $\P$;
    \STATE apply a Householder reflection $\Q_k$ to eliminate $r_{k+1,k}, r_{k+2,k}, \dots,r_{n,k}$;
    \STATE update $l_j$ by setting $l_j=l_j-r_{k,j}^2$, $j=k+1,\dots,m$;
  \ENDFOR
\end{algorithmic}
\end{algorithm}

Note that the cost of computation of $l_j$ in the algorithm is negligible compared with the other cost.
 
As Givens rotations have better numerical stability than GSO, 
in line \ref{l:pt} of Algorithm \ref{a:LLL}, we propose to use a Givens rotation to do triangularization.

\subsection{PLLL reduction algorithm}

Now we combine the strategies we proposed in the previous subsections
and give a description of the reduction process in Algorithm \ref{a:PLLL}, to be referred to as  a partial LLL (PLLL) reduction algorithm. 

\begin{algorithm}
 \caption{PLLL reduction}\label{a:PLLL}



\begin{algorithmic}[1]
  \STATE compute the Householder QR factorization with minimum pivoting: $\Q^T\H\P=\R$;
  \STATE set $\Z=\P$, $k=2$;
  \WHILE{$k\le n$}
   \STATE $\zeta=\lfloor r_{k-1,k}/r_{k-1,k-1} \rceil$, $\alpha=(r_{k-1,k}-\zeta r_{k-1,k-1})^2$;
   \IF{$\delta\, r_{k-1,k-1}^2>  (\alpha+r^2_{k,k})$}
    \IF{$\zeta\neq 0 $}
      \STATE apply the IGT $\Z_{k-1,k}$ to reduce $r_{k-1,k}$;
        \IF{$|\zeta|\geq 2$} 
            \FOR{$i=k-1,\dots,1$}
	        \STATE apply the IGT $\Z_{i,k}$ to reduce $r_{i,k}$;
           	\STATE update $\Z$:  $\Z=\Z\Z_{i,k}$;
           \ENDFOR
       \ENDIF 
      \ENDIF
      \STATE permute and triangularize: $\R=\G_{k-1,k}\R\P_{k-1,k}$;
    \STATE update  $\Z$:  $\Z=\Z\P_{k-1,k}$;
    \STATE $k=k-1$, when $k>2$;
   \ELSE
    \STATE $k=k+1$;
   \ENDIF
  \ENDWHILE
\end{algorithmic}
\end{algorithm}


\section{Numerical Experiments}\label{s:sim}
In this section we give numerical test results  to compare efficiency  and stability of LLL, ELLL and PLLL. 
Our simulations were performed in MATLAB 7.8 on a PC running Linux.
The parameter $\delta$ in the reduction was set to be $3/4$ in the experiments. 
Two types of matrices were tested. 
\begin{enumerate}
\item Type 1. The elements of $\H$ were drawn from an i.i.d. zero-mean, unit variance Gaussian distribution.
\item Type 2. $\H=\U\D\V^T$, where $\U$ and $\V$ are the Q-factors of the QR factorizations of random matrices
and $\D$ is a diagonal matrix, whose first half diagonal entries follow an i.i.d. uniform distribution over 10 to 100, 
and whose second half diagonal entries follow an i.i.d. uniform distribution over 0.1 to 1.
So the condition number of $\H$ is bounded up by 1000.
\end{enumerate}

\begin{figure}[!t]
\centering
\includegraphics[width=2.6in]{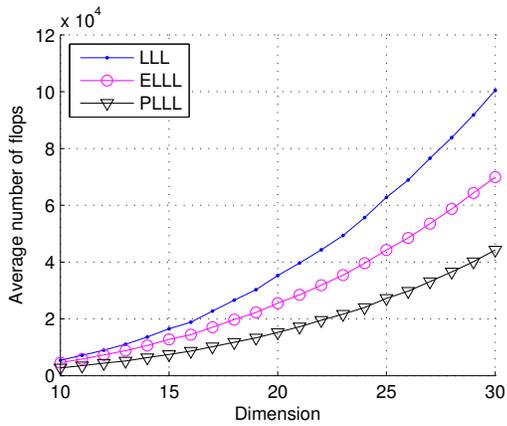}
\caption{Matrices of Type 1 - Flops}
\label{f:flpn}
\end{figure}

\begin{figure}[!t]
\centering
\includegraphics[width=2.6in]{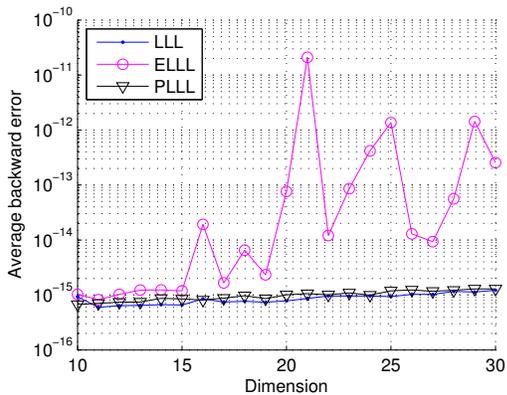}
\caption{Matrices of Type 1 - Backward Errors}
\label{f:errn}
\end{figure}

For matrices of Type 1, we gave 200 runs for each dimension $n$.
Figure \ref{f:flpn} gives the average flops of the three reduction algorithms,
and  Figure \ref{f:errn} gives the average relative backward error $\|\H-\Q_c\R_c\Z_c^{-1}\|_2/\|\H\|_2$,
where $\Q_c$, $\R_c$ and $\Z_c^{-1}$ are the computed factors of the QRZ factorization
produced by the reduction.
From Figure 1 we see that PLLL is faster than both LLL and ELLL. 
From Figure 2 we observe that  the relative backward error for both LLL and PLLL
behaves like $O(nu)$, where $u \approx 10^{-16}$ is the unit round off.
Thus the two algorithms are numerically stable for these matrices.
But ELLL is not numerically stable sometimes. 

\begin{figure}[!t]
\centering
\includegraphics[width=2.6in]{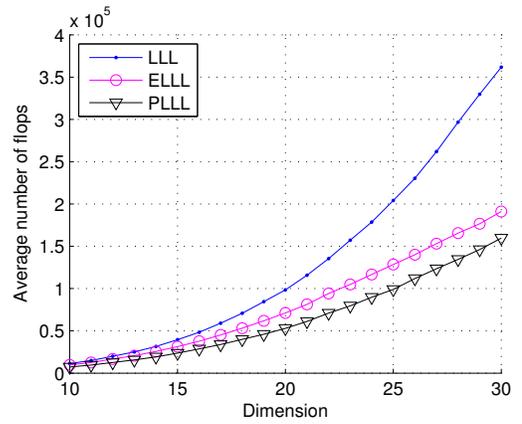}
\caption{Matrices of Type 2 - Flops}
\label{f:flpg}
\end{figure}

For matrices of Type 2, Figure \ref{f:flpg} displays the average flops of the three reduction algorithms over 200 runs for each dimension $n$. 
Again we see that PLLL is faster than both LLL and ELLL. 

\begin{figure}[!t]
\centering
\includegraphics[width=2.6in]{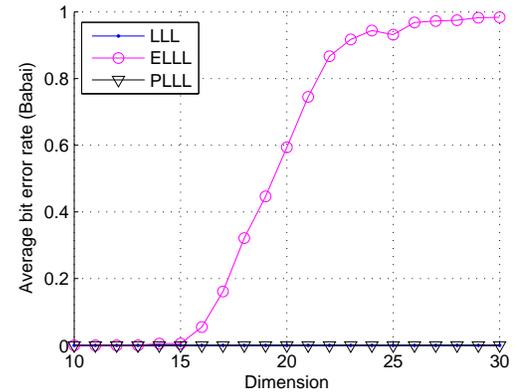}
\caption{Matrices of Type 2 - BER}
\label{f:berg}
\end{figure}

To see how the reduction affects the performance of the Babai integer point, for Type 2 of matrices,
we constructed the linear model $\y=\H\x+\v$, where $\x$ is an integer vector randomly generated and $\v\sim \mathcal{N}(0,0.2^2 \I)$. 
Figure \ref{f:berg} shows  the average bit error rate (BER) over 200 runs for  each dimension $n$. 
Form the results we observe that the computed Babai points obtained by using LLL and PLLL performed perfectly,
but the computed Babai points obtained by using ELLL performed badly when the dimension $n$ is larger than 15. 
Our simulations showed that the computed ILS solutions obtained by using the three reduction algorithms behaved similarly. 
All these indicate that  ELLL can give a very poor estimate of $\x$ due to its  numerical stability problem.

%

\bibliographystyle{IEEEtran}
\bibliography{IEEEabrv,./ILS}

\end{document}